\documentclass[conference,letterpaper]{IEEEtran}
\IEEEoverridecommandlockouts
\usepackage{graphicx}
\usepackage{color}
\usepackage{latexsym}
\usepackage{url}
\usepackage{amsmath}
\usepackage{booktabs}
\usepackage{multirow}
\usepackage{tablefootnote}
\usepackage{physics}
\usepackage{quantikz}
\usepackage{color}
\usepackage{listings}
\usepackage{quantikz}
\usepackage{ascmac}
\usepackage{fancybox}

\newcommand\copyrighttext{%
  \footnotesize \textcopyright 2024 IEEE. Personal use of this material is permitted.
  Permission from IEEE must be obtained for all other uses, in any current or future
  media, including reprinting/republishing this material for advertising or promotional
  purposes, creating new collective works, for resale or redistribution to servers or
  lists, or reuse of any copyrighted component of this work in other works.
  }
\newcommand\copyrightnotice{%
\begin{tikzpicture}[remember picture,overlay]
\node[anchor=south,yshift=10pt] at (current page.south) {\fbox{\parbox{\dimexpr\textwidth-\fboxsep-\fboxrule\relax}{\copyrighttext}}};
\end{tikzpicture}%
}

\def\Underline{\setbox0\hbox\bgroup\let\\\endUnderline}
\def\endUnderline{\vphantom{y}\egroup\smash{\underline{\box0}}\\}

\begin{document}

\title{Accelerating Decision Diagram-based Multi-node Quantum Simulation with Ring Communication and Automatic SWAP Insertion}

\author{\IEEEauthorblockN{
Yusuke Kimura\IEEEauthorrefmark{1}, Shaowen Li\IEEEauthorrefmark{2}, Hiroyuki Sato\IEEEauthorrefmark{2} and
Masahiro Fujita\IEEEauthorrefmark{3}}
\IEEEauthorblockA{\textit{Fujitsu Limited, Japan}\IEEEauthorrefmark{1}}
\IEEEauthorblockA{\textit{The University of Tokyo, Japan\IEEEauthorrefmark{2}\IEEEauthorrefmark{3}}\\
yusuke-kimura@fujitsu.com\IEEEauthorrefmark{1},
\{li-shaowen879,schuko\}@satolab.itc.u-tokyo.ac.jp\IEEEauthorrefmark{2},
fujita@ee.t.u-tokyo.ac.jp\IEEEauthorrefmark{3}}}

\maketitle
\copyrightnotice

\begin{abstract}
An N-bit quantum state requires a vector of length $2^N$, leading to an exponential increase in the required memory with N in conventional statevector-based quantum simulators. A proposed solution to this issue is the decision diagram-based quantum simulator, which can significantly decrease the necessary memory and is expected to operate faster for specific quantum circuits. However, decision diagram-based quantum simulators are not easily parallelizable because data must be manipulated dynamically, and most implementations run on one thread.
This paper introduces ring communication-based optimal parallelization and automatic swap insertion techniques for multi-node implementation of decision diagram-based quantum simulators. The ring communication approach is designed so that each node communicates with its neighboring nodes, which can facilitate faster and more parallel communication than broadcasting where one node needs to communicate with all nodes simultaneously. The automatic swap insertion method, an approach to minimize inter-node communication, has been employed in existing multi-node state vector-based simulators, but this paper proposes two methods specifically designed for decision diagram-based quantum simulators.
These techniques were implemented and evaluated using the Shor algorithm and random circuits with up to 38 qubits using a maximum of 256 nodes. The experimental results have revealed that multi-node implementation can reduce run-time by up to 26 times. For example, Shor circuits that need 38 qubits can finish simulation in 147 seconds. Additionally, it was shown that ring communication has a higher speed-up effect than broadcast communication, and the importance of selecting the appropriate automatic swap insertion method was revealed.
\end{abstract}


\section{Introduction}
Recent years have seen remarkable advancements in the development of quantum computers. Google has developed a superconducting quantum chip equipped with 53 quantum bits and claimed to have achieved quantum supremacy~\cite{Arute2019}. IBM has developed a quantum computer equipped with 433 quantum bits~\cite{ibm}, which is utilized for the research and development of various quantum algorithms. However, a limited number of researchers can access such large-scale quantum computers, and the cost of use is high. Moreover, current quantum computers are susceptible to noise, posing a challenge for algorithm development. Therefore, the research and development of quantum simulators that can mimic the operation of quantum computers is practically important for the development of quantum error correction and other quantum algorithms.

The most common type of quantum simulator is the statevector-based quantum simulator. An $N$-bit quantum state can be represented by a complex vector of length $2^N$, and the statevector-based simulators allocate memory for $2^N$ complex numbers. The widely-known simulators include IBM's Qiskit Aer~\cite{Qiskit} and Qulacs~\cite{Suzuki2021qulacsfast}. GPUs can also be utilized to perform matrix-vector manipulations quickly~\cite{bayraktar2023cuquantum}. Moreover, when complex numbers are represented using the double type, 16 GB is required to hold a 30-qubit statevector. Therefore, the limit is approximately 30 qubits in a personal computer for this type of simulations. Moreover, as the number of qubits increases, the size of the quantum circuit also increases, making run-time longer.

Several approaches have been proposed to solve these problems. The tensor network-based simulator holds the statevector as an MPS (Matrix Product State) \cite{Vidal_2003} and simulates a quantum circuit as a tensor network \cite{ORUS2014117}. It is particularly utilized in quantum machine learning \cite{McClean2018,Huggins_2019}. Although it has the potential to reduce the required memory and speed up the simulation, it has the drawback of slowing down operations for quantum states with many entanglements or deep circuits. Another type of simulator is the decision diagram-based simulator. Decision diagrams have been used as a method for manipulating matrices, such as inverse matrix calculations, to represent data structures such as logical functions and matrices~\cite{10.1109/TC.1986.1676819,Fujita1997}. It is also possible to use them to store and manipulate quantum states (vectors) and quantum circuits (matrices)~\cite{QMDD}. Details will be described later, but it is possible to reduce the amount of memory and accelerate the simulation when there are common parts in the subvector or when there are few values other than zero (in sparse cases). The well-known implementations include DDSIM~\cite{ddsim} and SliQSim~\cite{sliqsim}.

Another orthogonal approach is multi-node implementation. In a multi-node environment, multiple independent computers are connected via optical fiber or similar means to realize comulti-node communications. Currently, it is common to use a standardized communication method called MPI. By implementing on multi-node machines, it is possible to collectively utilize memory space spanning over multiple nodes, and more quantum bits can be handled. Additionally, because calculations are distributed and parallelized, acceleration can be expected. In mpiQulacs~\cite{imamura2022mpiqulacs} where the statevector-based quantum simulator Qulacs~\cite{Suzuki2021qulacsfast} is multi-noded, a simulation of 36 quantum bits is performed with 64 nodes. Similar research includes Intel-QS~\cite{Guerreschi_2020}. A method of compressing the statevector can possibly be combined for a specific algorithm and \cite{10.1109/TPDS.2019.2947511} simulated a 49-qubit Quantum Supremacy circuit on the Sunway TaihuLight supercomputer. Statevector-based quantum simulators can also perform simulations using GPUs. In \cite{9259953}, 16 nodes equipped with six NVIDIA Tesla V100 were used, and a maximum of 34-qubit simulation was performed while communicating via MPI. \cite{Shang2023} introduced an MPS simulator with MPI, which can simulate a chemical simulation (H2-Chain) of up to 1000 quantum bits on an eight-node environment. As described above, there are several existing studies on multi-node implementation in statevector- and tensor network (MPS)-based simulators.

The main contribution of this study is to propose multiple methods (ring communication and automatic swap insertion method) to efficiently implement a multi-node decision diagram-based quantum simulator on multi-node machines. To the best of the authors' knowledge, this paper is the first to introduce the multi-node decision diagram-based quantum simulator. In the experiments, evaluations were conducted with the QCBM (random) circuit and Shor algorithm, and results showed the execution time was shortened by up to approximately 26 times. Furthermore, the run-time difference between multi-node decision diagram-based and statevector-based quantum simulators, i.e., which simulator behaves better in which circuits, was clarified.

The structure of this paper is as follows. In Sec. \ref{sec:background}, the mechanism of the decision diagram-based simulator and the multi-node implementation method of the existing statevector-based simulator are explained. In Sec. \ref{sec:methods}, methods for executing the decision diagram-based simulator on multi-node, including details such as communication methods and automatic swap insertion methods, are explained. Sec. \ref{sec:experiments} and \ref{sec:evaluation} explain the experimental environment and show the effects of the proposed method and the comparison with the statevector-based simulator (Qulacs). The paper is summarized in Sec. \ref{sec:summary}, and future directions are also discussed.

\section{Background Knowledge}\label{sec:background}
\subsection{Decision diagram-based quantum simulation}
This section introduces the decision diagram employed for the representation of quantum states and quantum gates. Although existing research utilizes several types of decision diagrams, this study uses QMDD\cite{QMDD} adopted by DDSIM\cite{ddsim}. Owing to page limit constraints, measurement gates and other details are not addressed, so please refer to the existing literature \cite{QMDD,ddsim}.

In this paper, nodes in the decision diagram are called “graph nodes.” Additionally, computational nodes in a multi-node environment are simply called “nodes” to distinguish from graph nodes.

\subsubsection{Representation of quantum state (vector)}\label{sec:background:sv}
\begin{figure}[tb]
    \centering
    \includegraphics[width=50mm]{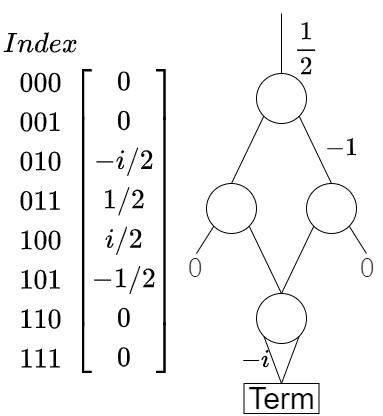}
    \caption{Statevector in a decision diagram}
    \label{fig:dd:vector}
\end{figure}

First, we explain how to represent quantum states in decision diagrams. A statevector of N qubits is a complex vector of length $2^N$. For instance, consider the vector depicted on the left side of Fig. \ref{fig:dd:vector}. This vector can also be represented by a decision diagram shown on the right. "Term" stands for a terminal graph node. To determine the value of each element in the vector from the decision diagram, we can trace the edges according to the value of the index and calculate the product of the edge weights. Let index value 0 go left and index value 1 go right, and let the edge value for an edge without specified value be 1. For example, if you want a value with an index of 101, follow the edges from the top in the order of right, left, and right. Because the edge weights are $(\frac{1}{2},-1,1, 1)$, the product is $-\frac{1}{2}$. If the index is 110, an edge weight of $0$ appears in the middle; therefore, the product is $0$.

Please note that the original vector comprises eight complex numbers, whereas the decision diagram contains only four graph nodes. Thus, when the same subvector appears, the graph nodes can be shared on the decision diagram, potentially leading to significant memory savings. Furthermore, the number of graph nodes can be reduced when there are zeros in the vector.

\subsubsection{Representation of quantum gate (matrix)}\label{sec:background:mat}
\begin{figure}[tb]
    \centering
    \includegraphics[width=80mm]{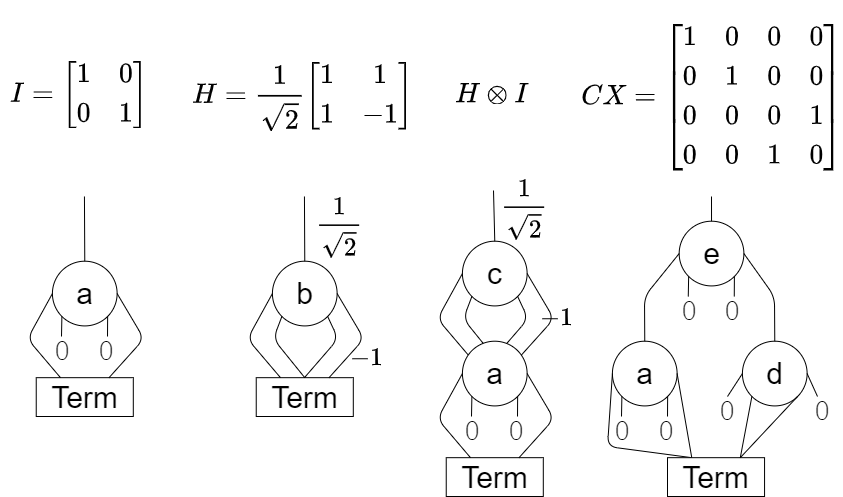}
    \caption{Quantum gate in decision diagrams}
    \label{fig:dd:matrix}
\end{figure}

Next, how to represent quantum gates (in matrix) in decision diagrams is explained. Quantum gates that act on N quantum bits can be represented by a square matrix of size $2^{N}$. While binary graphs were used to represent vectors, one graph node has four sub-nodes in the case of a square matrix (or, two splits can be repeated twice). To obtain the decision diagram of a two-qubit gate from two 1-qubit gates, the two decision diagrams can be simply combined. When two quantum bits are entangled like a CNOT gate, a specific decision diagram that matches the unitary matrix is required. Some examples are shown in Fig. \ref{fig:dd:matrix}.

As with vectors, graph nodes can be shared in matrices as well. Therefore, a memory reduction effect can be expected when the quantum gate is expressed in a decision diagram. For instance, a 2-qubit quantum gate $H\bigotimes I $ requires 16 complex numbers for the unitary matrix. In the case of a decision diagram, it can be represented by only two graph nodes and nine weights, including the root edge.

The matrix-vector product in the form of a decision diagram can be recursively computed from the root graph nodes. Code \ref{code:mul} provides an overview of the multiplication, and more detail can be found in the existing research \cite{QMDD, ddsim}.

\begin{figure}[tb]
\begin{lstlisting}[caption=Multiplication, label=code:mul, numbers=none]
def mul(m: Edge, v: Edge):
  weight = m.weight*v.weight
  if isTerminal(m) and isTermainl(v):
    node = Terminal
  left = add(mul(m.child(0),v.child(0)),
             mul(m.child(1),v.child(1)))
  right =add(mul(m.child(2),v.child(0)),
             mul(m.child(3),v.child(1)))
  node = Node(left, right)
  return Edge(weight, node)
\end{lstlisting}
\end{figure}

\subsection{Statevector-based multi-node quantum simulation}\label{sec:background:sv_multi-node}
\begin{figure}[tb]
    \centering
    \includegraphics[width=50mm]{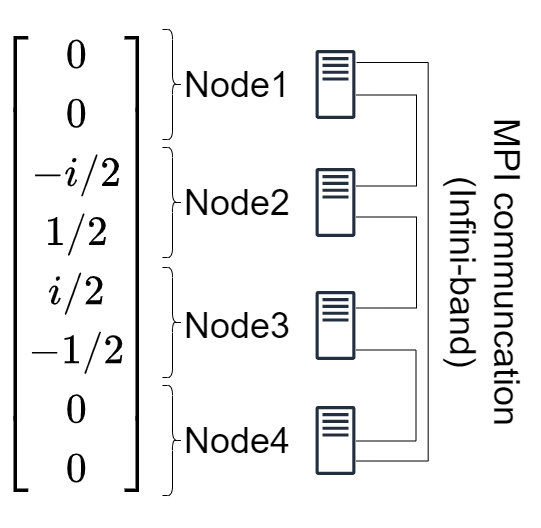}
    \caption{Statevector distribution}
    \label{fig:sv:multi}
\end{figure}

Statevectors are equally divided and placed on multiple nodes. For example, if a 3-bit quantum state is managed by four nodes, as shown in Fig. \ref{fig:sv:multi}, each node is responsible for two elements. MPI is used for communication between nodes, and the InfiniBand is often used for physical connections.
Details can be found in the GitHub repository~\footnote {Intel-QS: \url{https://github.com/intel/intel-qs}}~\footnote{Qulacs: \url{https://github.com/qulacs/qulacs}} and papers~\cite{imamura2022mpiqulacs, Guerreschi_2020}.

\subsection{SWAP Insertion}\label{sec:background:swap}
\begin{figure}[tb]
    \centering
    \includegraphics[width=80mm]{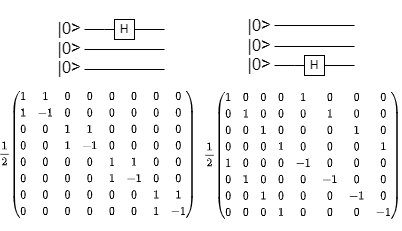}
    \caption{Unitary matrix of two gates}
    \label{fig:swap:unitary}
\end{figure}

In the simulator discussed in the previous section, the amount of MPI communication varies depending on the qubit position where quantum gates are applied. For example, Fig.~\ref{fig:swap:unitary} shows unitary matrices of two quantum gates in a three-qubit quantum circuit, in which one H gate is applied to the first bit in the left and the other H gate is applied to the third bit in the right. When the statevector is stored across multiple nodes as illustrated in Fig.~\ref{fig:sv:multi}, the calculation of the left gate can be operated within each node, while the calculation of the right gate requires communication among nodes. More generally, when $2^M$ nodes are used to simulate a $N$ qubits circuit, a quantum gate applied to the first $N-M$ qubits only requires calculation within each node, whereas a gate applied on the last $M$ qubits requires inter-node communication. These are referred to as local and global areas, respectively. With an increase of gates operating in the global area, MPI communication also increases.

\begin{figure}[tb]
    \centering
    \includegraphics[width=60mm]{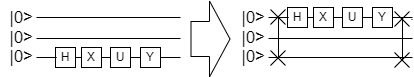}
    \caption{SWAP gate \& MPI communication}
    \label{fig:swap:auto}
\end{figure}

Therefore, it is preferable to apply gates to the local area. To satisfy this requirement, mpiQulacs~\cite{imamura2022mpiqulacs} and the distributed GPU version of Qiskit Aer~\cite{9259953} implemented a technique to reduce MPI communication by automatically inserting swap gates. In the left circuit of Fig.~\ref{fig:swap:auto}, four gates applied to the last bit, resulting in four MPI communications. However, if a swap gate is inserted before and after the gates, the resulting statevector is exactly the same, but MPI communication is only required for the two swap gates. Although the required computation increases, the overall execution time is reduced owing to the decrease in inter-node communication. This automatic insertion of the swap gate accelerates the simulation by up to 10 times \cite{9259953}.

\section{Multi-node Quantum Simulation with Decision Diagram}\label{sec:methods}
\subsection{Distribution of statevector}
\begin{figure}[tb]
    \centering
    \includegraphics[width=60mm]{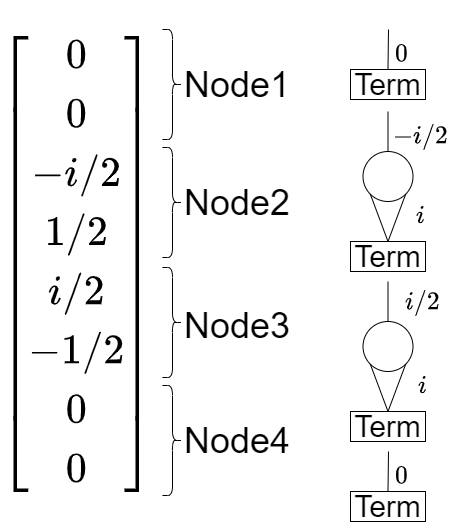}
    \caption{Statevector distribution with decision diagrams}
    \label{fig:dd:multi}
\end{figure}

This section explains how the statevector is stored when a decision diagram-based quantum simulator is used with a multi-node environment. As explained in Sec. \ref{sec:background:sv}, a decision diagram-based quantum simulator holds the statevector in the form of a decision diagram. In this study, when multiple nodes are used, the statevector is divided equally and stored in the form of a decision diagram in each node. Fig. \ref{fig:dd:multi} shows how to divide a three-qubit statevector into four nodes using decision diagrams. Compared to Fig. \ref{fig:dd:vector}, the decision diagram of each node is smaller, which is expected to reduce memory usage and computation time.

In this study, we decided that the number of nodes is limited to a power of two ($2^M=1, 2, 4, 8, 16,\dots$). Moreover, the number of nodes $2^M$ must be the same or smaller than $2^N$, where $N$ is the number of qubits. There is no necessity that the number of nodes is $2^M$, but the vector length $2^N$ can be equally divided with $2^M$ sub-vectors, making it easier to implement. Specifically, it is easier to extract the sub-matrix of the quantum gate and to distinguish the local and global areas in Sec. \ref{sec:background:swap}.

If the above method is adopted, the size of the decision diagram between nodes can be significantly unbalanced.
It is possible to change the division boundary dynamically, but this is not implemented in this study.

\subsection{MPI communication}\label{sec:methods:comm}
In this section, we explain how quantum gates are applied to the distributed quantum states. In a quantum simulation, a matrix-vector (quantum gate-statevector) product is calculated. The multiplication with four nodes can be shown as eq. (\ref{eq:mul}), where the sub-vectors of individual nodes are $v1, v2, v3, v4$. Note that $w11, \dots, w44$ are sub-matrices of a quantum gate. A unitary matrix or a sub-matrix of a quantum gate is based on a limited number of basis gates and can be easily generated in each node.

\begin{equation}
\begin{split}
\left(\begin{array}{c}v'1\\v'2\\v'3\\v'4\end{array}\right)=\left(\begin{array}{cccc}w11&w12&w13&w14\\w21&w22&w23&w24\\w31&w32&w33&w34\\w41&w42&w43&w44\end{array}\right)\left(\begin{array}{c}v1\\v2\\v3\\v4\end{array}\right)\\
=\left(\begin{array}{c}
w11\times v1+w12\times v2+w13\times v3+w14\times v4\\
w21\times v1+w22\times v2+w23\times v3+w24\times v4\\
w31\times v1+w32\times v2+w33\times v3+w34\times v4\\
w41\times v1+w42\times v2+w43\times v3+w44\times v4\end{array}\right)\label{eq:mul}
\end{split}
\end{equation}

\begin{figure}[tb]
    \centering
    \includegraphics[width=75mm]{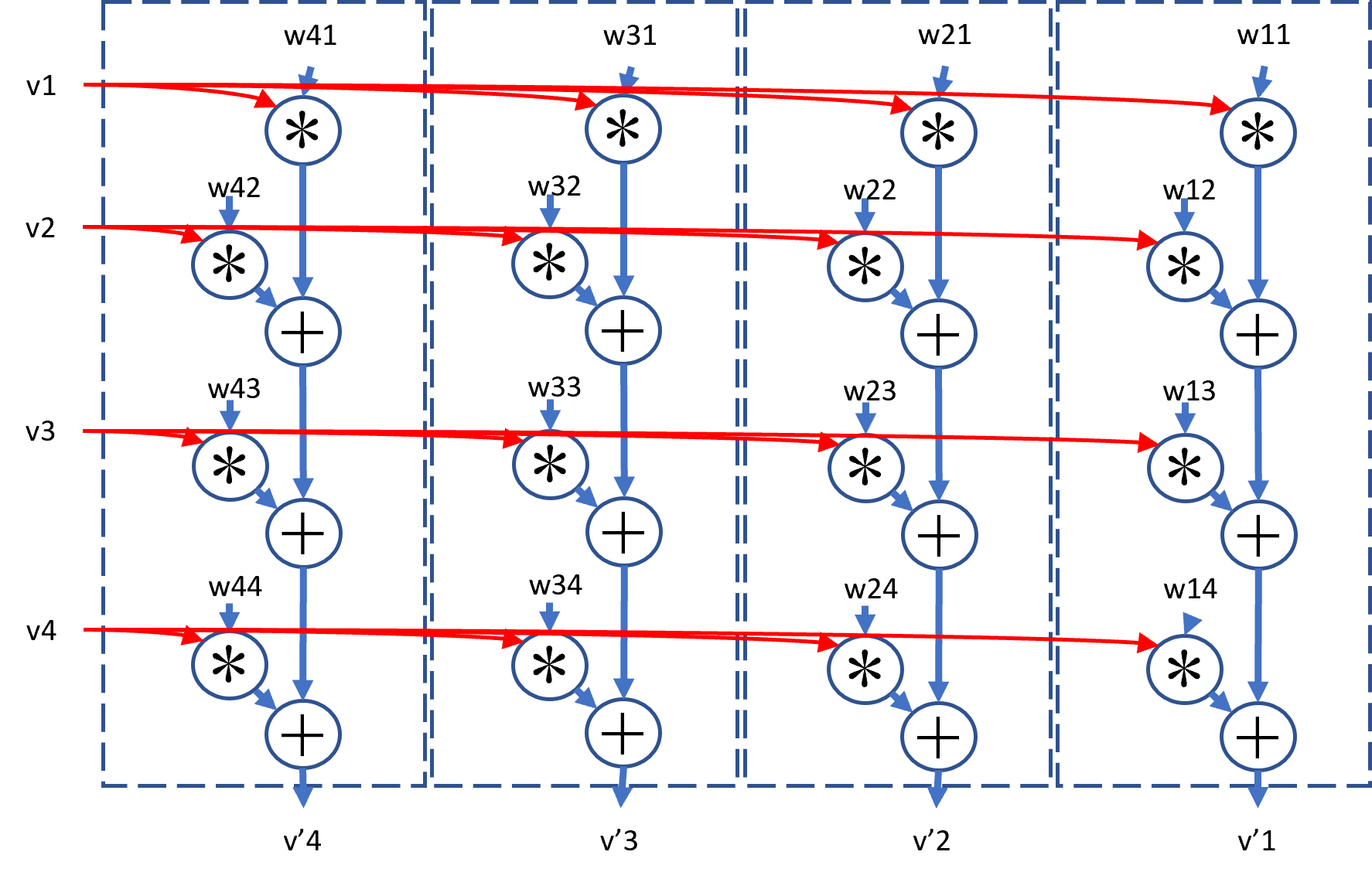}
    \caption{MPI communication with broadcast}
    \label{fig:mpi:bcast}
    \includegraphics[width=70mm]{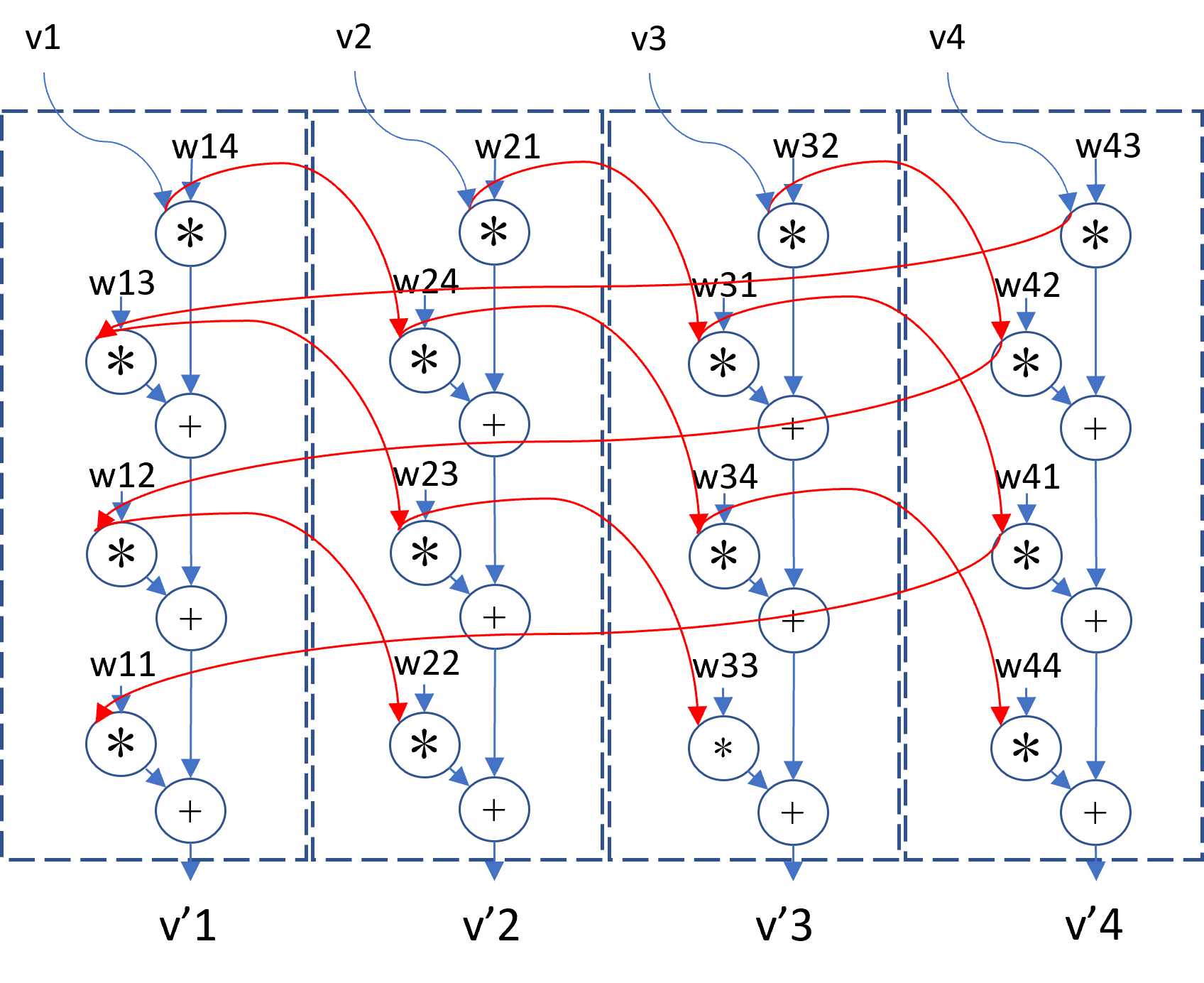}
    \caption{MPI communication with neighbors (ring)\\The order of calculations is different from the one in Fig. \ref{fig:mpi:bcast}, although the final results are the same.}
    \label{fig:mpi:ring}
\end{figure}

As shown in eq. (\ref{eq:mul}), $v1$ is required when calculating all of $v'2, v'3, v'4$; therefore, so $v1$ must be delivered to each node.
The most intuitive way for humans is that each node $i$ broadcasts its sub-vector $v_i$ to all nodes, as illustrated in Fig. \ref{fig:mpi:bcast}.

In another method, it is assumed that nodes are connected in a ring, where each node communicates only with its neighbors and all nodes can communicate in parallel.
$v1$ is first transmitted from the first node to the second node, and the second node distributes it to the third node and so on.
This bucket relay transmission is illustrated in Fig. \ref{fig:mpi:ring}. This is using a different order of calculations from Fig. \ref{fig:mpi:bcast}, but the result is the same.

Because the final calculation results are the same among both methods, only the communication method differs.
Broadcast communications, in which one node distributes data to all other nodes, are likely to have larger communication overhead than ring communications, in which only adjacent nodes communicate.
The experimental results are shown in Sec. \ref{sec:experiments}, and it is found that the ring communication is approximately 10\%-20\% faster in the Shor algorithm. Additionally, the effect is larger when the number of nodes is larger. In the random circuit, the ring communication performs approximately six times faster than broadcast communication.

\subsection{Auto swap gates insertion}\label{sec:methods:swap}
As discussed in Sec. \ref{sec:background:swap}, multi-node quantum simulations can leverage swap gates to reduce MPI communication and speed up computations.
This section describes our implementation of the auto swap insertion.

First, we decided to insert the swap gate automatically when a gate is applied to the global area.
At this time, it is important which qubits are moved from the local area to the global area.
Qubits placed in the global should not be affected by later quantum gates as much as possible.
Therefore, we decided to analyze the quantum gates in order and insert the affected qubits into a list that should be in the local area.
When the number of qubits in the list exceeds the size of the local area ($N-M$), the remaining qubits are placed in the global area.
The above process is written in Code \ref{code:swap_step1}. This gives the next list of qubits (next\_global) that should be in the global area.

\begin{figure}[tb]
\begin{lstlisting}[caption=Extracting global qubits, label=code:swap_step1, numbers=none]
def do_swap(current_gate):
  nTotal=N, nGlobal=M, nLocal=N-M
  all_bits = {1, 2, ..., N}
  next_local = set()
  
  while current_gate.hasNext():
    candidates = current_gate.get_applied_qubits()
    for a qubit in candidates:
      next_local.insert(qubit)
      if len(localBist)==nLocal:
        break
    current_gate = current_gate.next()

  next_global = all_bits - next_local
\end{lstlisting}
\end{figure}

By the above process, two lists are obtained: the list of qubits that are now in the global area (current\_global) and the list of qubits that will be in the global area the next time (next\_global).
SWAP gates are automatically inserted based on these lists. In this study, we propose the following two automatic insertion methods.
\begin{itemize}
\item v1: Insertion method that preserves the original qubit order as much as possible
\item v2: Insertion method with a minimum number of swap gates
\end{itemize}

v1 is a method specific to the decision diagram-based quantum simulator that assumes that the original order works relatively well. The shape of the decision diagram can vary significantly depending on the bit order, resulting in an increase or decrease in execution time. The advantage of this method is that it is designed to maintain the original order of qubits in both the local and global area so that there is little chance of falling into an unacceptable bit order.
v2 is a method that does not consider the original qubit order. In a random circuit where statevector values vary, changing the qubit order may not significantly change the shape of the decision diagram. In this case, v2 is expected to run faster.

\subsubsection{v1}
In this method, the original qubit order is maintained in the local area and the global area.
The current qubit order is compared with the next qubit order from the beginning, and if they are different, a swap gate is inserted so that they become the same.
It has a disadvantage in the number of swaps compared to v2, but the advantage is that the original qubit order is preserved as much as possible.

For example, consider a case where the original order is $[a,b,c,d,e]$, the global area is 1-qubit long, and $c$ is moved to the global area (the last place). In this case, the next order should be $[a,b,d,e,c]$, which will be realized by two swap gates as follows.
\begin{eqnarray}
[a,b,c,d,e] &\to& SWAP(3,4) \to [a,b,d,c,e] \nonumber\\
            &\to& SWAP(4,5) \to [a,b,d,e,c] \nonumber
\end{eqnarray}

\subsubsection{v2}
From current\_global and current\_global, the lists of qubits that should be moved from global to local and from local to global can be obtained. The qubits should be swapped one by one with a swap gate. This method has the advantage of requiring fewer swaps than v1. However, it is important to note that the qubit order can become different from the original one.

Consider the same example in 1). Because v2 does not consider the order of the original qubits, it uses only one swap gate, resulting in $[a, b, e, d, c]$. This gives a different ordering than v1.
\begin{eqnarray}
[a,b,c,d,e] \to SWAP(3,5) \to [a,b,e,d,c] \nonumber
\end{eqnarray}

\section{Experimental Results}\label{sec:experiments}
\subsection{Environment}
We have developed a decision diagram-based quantum simulator using QMDD \cite{QMDD}. The implementation was based on DDSIM \cite{ddsim}, but fully customized for the multi-node functionality. We will disclose our GitHub URL after the double-blind review process. The main portion of the simulator is implemented in 3,000 lines of C++. It is also designed to work as a Qiskit Backend, which is implemented in 1,000 lines of Python. For MPI communication, the Fujitsu MPI library was used via Boost-MPI. Fujitsu MPI is a specialized library for the hardware environment described below. Note that the programs in each node run in single-threaded.

A multi-node statevector-based quantum simulator was also prepared for comparison. Qulacs v0.6.2 was used \cite{Suzuki2021qulacsfast,imamura2022mpiqulacs}~\footnote{Qulacs: \url{https://github.com/qulacs/qulacs}}, and we followed the build instructions to enable MPI functionality. The same compilers and libraries were used.

We used Wisteria-O provided by the Information Technology Center of the University of Tokyo. Please refer to the website \footnote{Wisteria-O: \url{www.cc.u-tokyo.ac.jp/en/supercomputer/wisteria/system.php}} for details. This system has 7,680 nodes, but this study used a maximum of 256 nodes owing to budget and waiting time.

\begin{itemize}
\item CPU: Fujitsu A64FX Arm Processor (48 cores, 2.2GHz) 
\item Memory: 32GiB HBM2 memory (1024 GB/s)
\item Interconnect: Tofu InterConnect-D (28Gbps $\times$ 2lane $\times$ 10port)
\item OS: RHEL8.3, Kernel 4.18
\item Software: GCC 8.3.1, Python 3.8.9
\item Libraries: Fujitsu-MPI 1.2.38, Boost 1.81.0
\end{itemize}

\subsection{Experiment 1. Shor}\label{sec:shor}
Shor algorithm~\cite{Shor} is used for factorization (e.g., 253 can be factorized to $23\times 11$). The quantum circuit for the Shor algorithm used in this experiment requires $4n+2$ qubits, where $n$ is the bit length of the number to be factored. For example, factoring 15 (1111, $n=4$) requires 18 qubits. Existing research suggests methods using $2n+3$~\cite{shor_2n_3} or $2n+2$~\cite{shor_2n_2} qubits. These algorithms should work correctly, but in this study, we used $4n+2$ due to code availability.

The experimental results are shown in Tab.~\ref{tab:shor}. DD shows the run-time of the decision diagram-based quantum simulator, and SV is the run-time of the statevector-based simulator (Qulacs). We investigated how the execution time varies depending on the combination of the communication method (Ring or Bcast) and the presence or absence of automatic swap insertion (No, v1, v2). The number of nodes used is 1-256 nodes. When the number of nodes is 1, the execution time is the same for all conditions because the communication method and the automatic swap insertion do not affect the execution.

In the case of the Shor algorithm, the statevector-based simulator could not factor 511 because 38 qubits require a large amount of memory, and the statevector-based simulator became out of memory error. Therefore, we decided to show the result of factoring 57 for statevector-based. Accordingly, we conducted experiments with both 26 and 38 qubits for the decision diagram-based simulator. Additionally, please refer to \cite{imamura2022mpiqulacs} about MPI communication and auto swap gate insertion of Qulacs.

Fig.~\ref{fig:results:shor} is the line graph to help understand the relationship between the number of nodes and the relative execution time. The horizontal axis is the number of nodes, and the vertical axis is the run-time where 1 is the result of one node. The graphs used 38 qubit results for the decision diagram-based and 26 qubit results for the statevector-based simulator. Fig.~\ref{fig:results:shor:nosv} is the line graph without statevector-based results, and its vertical axis is not logged.

\begin{table*}[t]
\centering
\caption{Shor Experimental Results (DD:Decision Diagram-based, SV: Statevector-based)}
\label{tab:shor}
\begin{tabular}{@{}crrrccrrrrrrrrr@{}}
\toprule
\multicolumn{1}{c}{\multirow{2}{*}{Simulator}} & \multicolumn{1}{c}{\multirow{2}{*}{Number}} & \multicolumn{1}{c}{\multirow{2}{*}{nQubits}} & \multicolumn{1}{c}{\multirow{2}{*}{nGates}} & \multicolumn{1}{c}{\multirow{2}{*}{MPI}} & \multicolumn{1}{c}{\multirow{2}{*}{SWAP}} & \multicolumn{9}{c}{nNodes \& run-time (sec)} \\ \cmidrule(l){7-15} 
\multicolumn{1}{c}{} & \multicolumn{1}{c}{} & \multicolumn{1}{c}{} & \multicolumn{1}{c}{} & \multicolumn{1}{c}{} & \multicolumn{1}{c}{} & 1 & 2 & 4 & 8 & 16 & 32 & 64 & 128 & 256 \\ \midrule
DD & 511 & 38 & 400,123 & Ring & No & 3,881 & 2,225 & 1,954 & 1,307 & 803 & 701 & 719 & 800 & 1,150 \\
DD & 511 & 38 & 400,123 & Ring & v1 & 3,881 & 1,115 & 1,106 & 461 & 393 & 289 & 233 & 177 & 147 \\
DD & 511 & 38 & 400,123 & Ring & v2 & 3,881 & 1,251 & 1,157 & 502 & 429 & 319 & 272 & 225 & 239 \\
DD & 511 & 38 & 400,123 & Bcast & No & 3,881 & 2,363 & 2,062 & 1,198 & 870 & 924 & 1,129 & 1,681 & 3,225 \\
DD & 511 & 38 & 400,123 & Bcast & v1 & 3,881 & 1,186 & 1,179 & 490 & 424 & 312 & 260 & 206 & 197 \\
DD & 511 & 38 & 400,123 & Bcast & v2 & 3,881 & 1,259 & 1,160 & 508 & 432 & 321 & 293 & 266 & 345 \\
DD & 57 & 26 & 51,123 & Ring & No & 40 & 38 & 30 & 25 & 27 & 34 & 53 & 91 & 170\\
DD & 57 & 26 & 51,123 & Ring & v1 & 39 & 30 & 21 & 17 & 14 & 14 & 15 & 16 & 24\\
DD & 57 & 26 & 51,123 & Ring & v2 & 39 & 31 & 23 & 19 & 16 & 16 & 18 & 28 & 60\\
DD & 57 & 26 & 51,123 & Bcast & No & 39 & 39 & 32 & 31 & 38 & 56 & 99 & 197 & 441\\
DD & 57 & 26 & 51,123 & Bcast & v1 & 39 & 30 & 22 & 17 & 16 & 15 & 17 & 31 & 79\\
DD & 57 & 26 & 51123 & Bcast & v2 & 39 & 32 & 24 & 19 & 17 & 18 & 23 & 45 & 119\\
SV & 57 & 26 & 51,123 & \cite{imamura2022mpiqulacs} & No & 48,039 & 25,928 & 14,509 & 8,634 & 5,539 & 4,043 & 2,230 & 1,435 & 1,150 \\
SV & 57 & 26 & 51,123 & \cite{imamura2022mpiqulacs} & Yes & 48,039 & 24,714 & 12,706 & 7,242 & 3,705 & 2,097 & 1,315 & 845 & 548 \\ \bottomrule
\end{tabular}
\end{table*}

\subsection{Experiment 2. QCBM}\label{sec:qcbm}
\begin{figure}[tb]
\centering
\begin{quantikz}
    &\gate{R_Z(\theta_1)}& \gate{R_X(\theta_2)}& \gate{R_Z(\theta_3)}& \ctrl{1} & \qw      & \targ{}  & \\
    &\gate{R_Z(\theta_1)}& \gate{R_X(\theta_2)}& \gate{R_Z(\theta_3)}& \targ{}  & \ctrl{1} & \qw      & \\
    &\gate{R_Z(\theta_1)}& \gate{R_X(\theta_2)}& \gate{R_Z(\theta_3)}& \qw      & \targ{}  & \ctrl{-2}& \\
\end{quantikz}
\caption{QCBM (Iterative portion in case of 3 qubits)}
\label{fig:qcbm}
\end{figure}
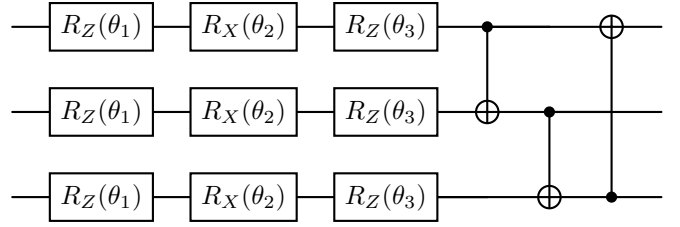

QCBM is a random circuit used in the evaluation of Qulacs \cite{Suzuki2021qulacsfast}. QCBM is an example in which the sharing of graph nodes is difficult in decision diagrams.
A sub-circuit consisting of RX, RZ, and CX gates with random parameters is shown in Fig. \ref{fig:qcbm}. It is repeated eight times.

The experimental results are shown in Tab. \ref{tab:qcbm}. The column information is similar to Tab. \ref{tab:shor}.
In the case of QCBM, the decision diagram-based simulator had a longer run-time compared to the statevector-based simulator and could not simulate 29-bit circuits in a realistic time. Therefore, the decision diagram-based uses the result of the 20-bit circuit. The statevector-based simulator performed both 20 and 29 qubits experiments.

\begin{figure}[ht]
  \centering
  \begin{minipage}[b]{0.48\textwidth}
  \includegraphics[width=\columnwidth]{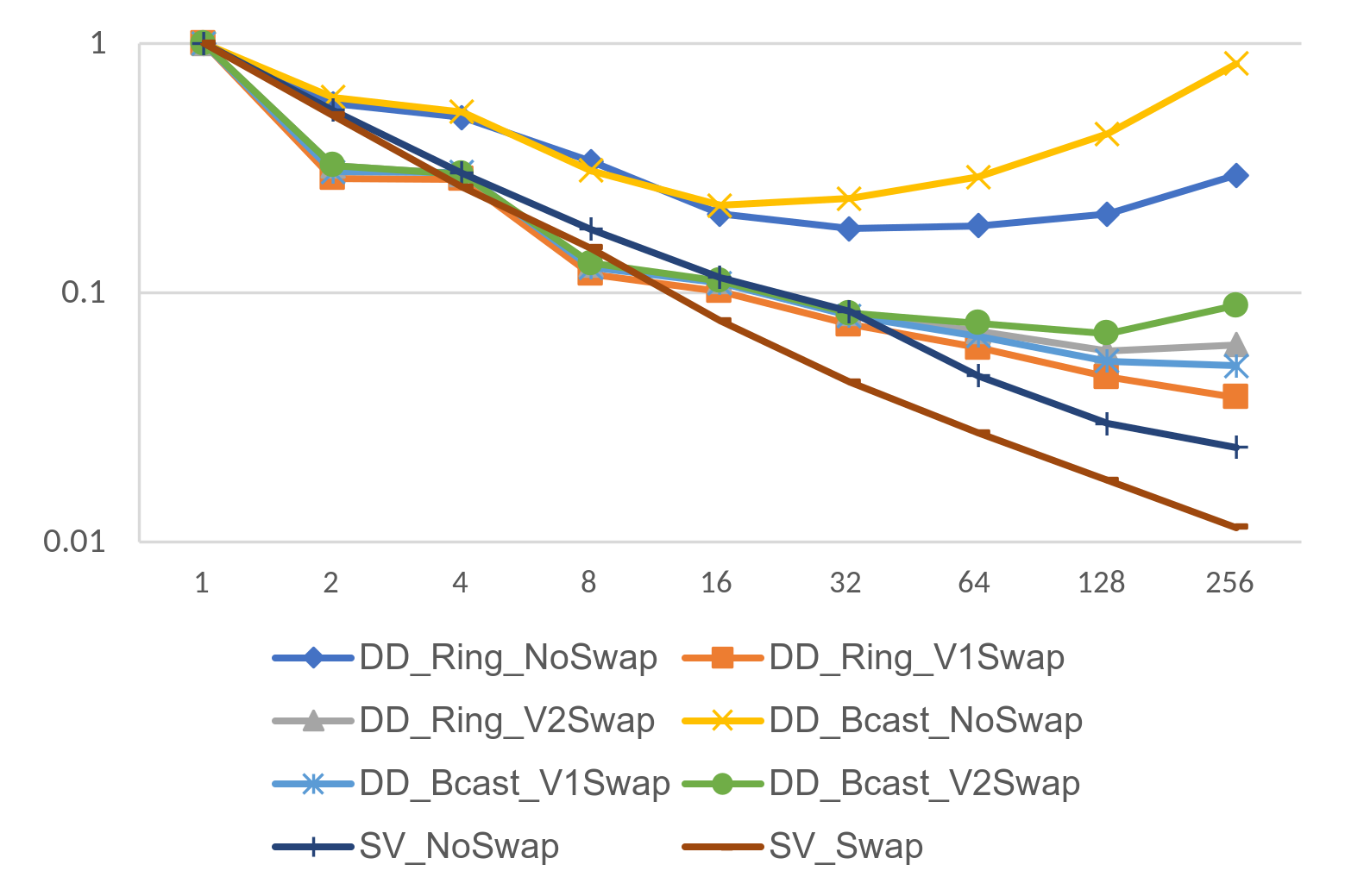}
  \caption{Shor: nNodes and run-time (the run-time of one node is 1)\\DD: 511 (38 qubits), SV: 57 (26 qubits)}
  \label{fig:results:shor}
  \end{minipage}
  \begin{minipage}[b]{0.48\textwidth}
  \includegraphics[width=\columnwidth]{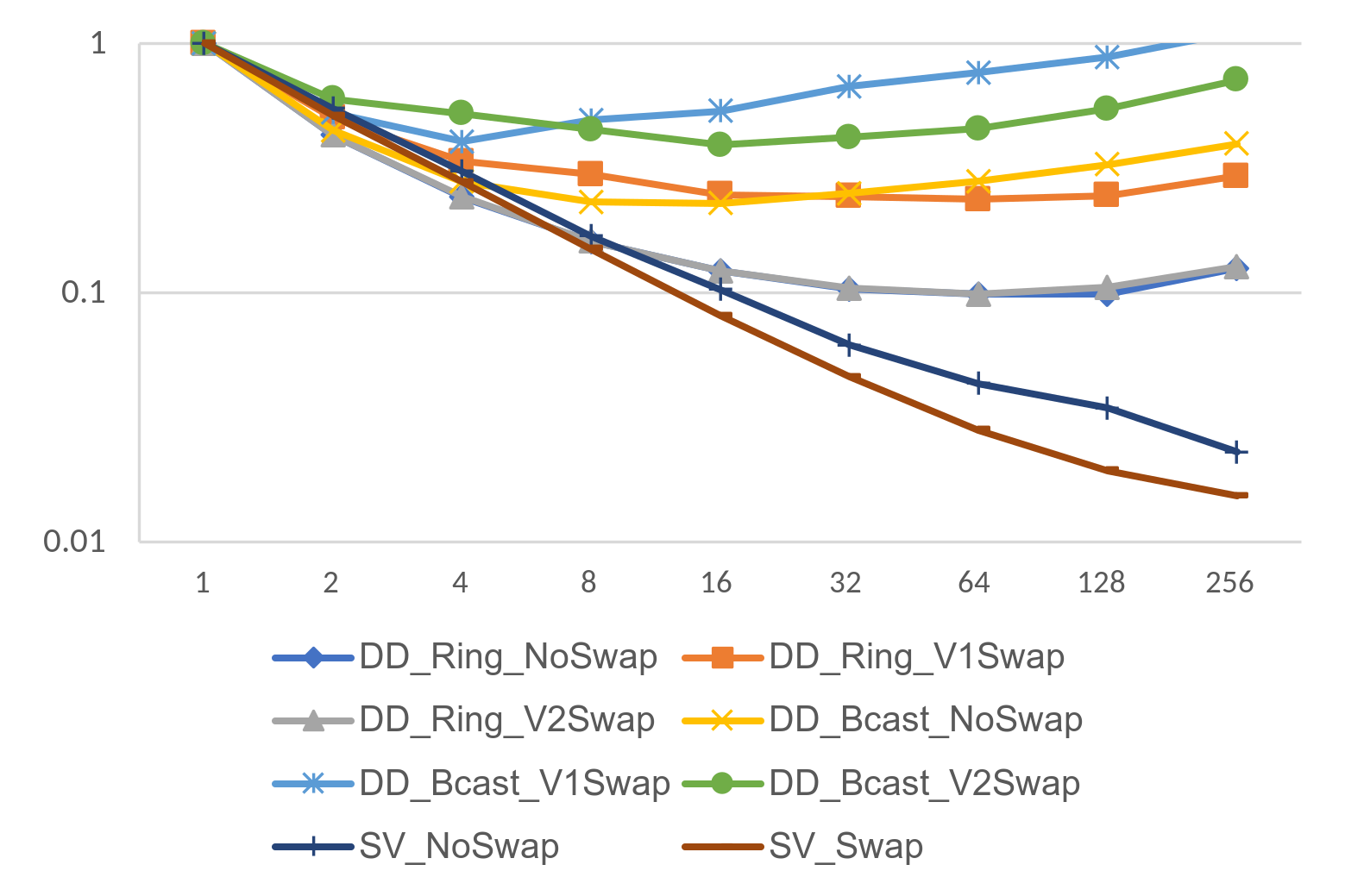}
  \caption{QCBM: nNodes and run-time (the run-time of one node is 1)\\DD: 20 qubits, SV: 29 qubits\\DD\_Ring\_NoSwap and DD\_Ring\_V2Swap overlap.}
  \label{fig:results:qcbm}
  \end{minipage}
\end{figure}

\begin{figure*}[ht]
  \centering
  \begin{minipage}[b]{0.48\textwidth}
  \includegraphics[width=\columnwidth]{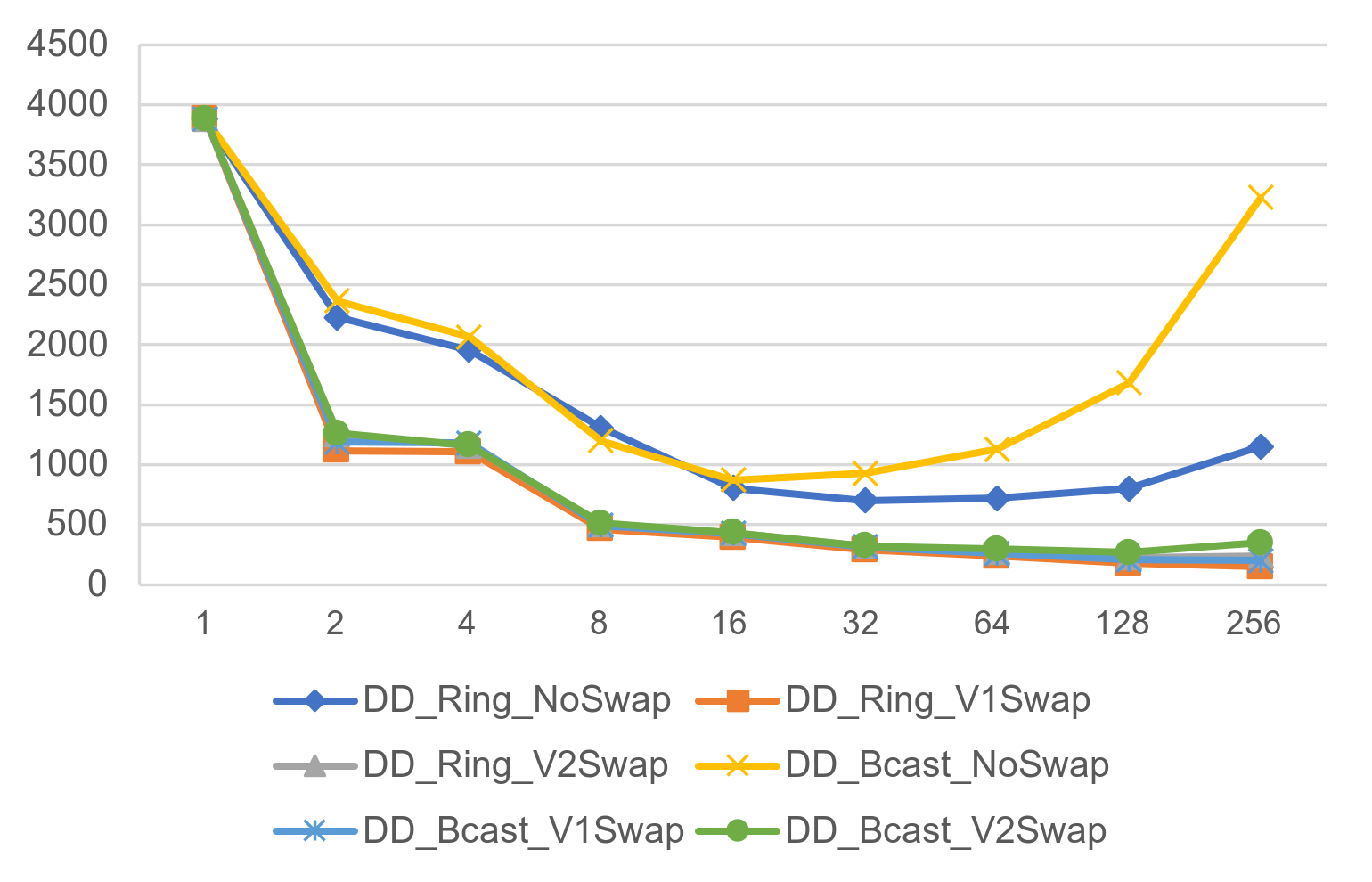}
  \caption{Shor: nNodes and run-time (sec) without statevector-based results}
  \label{fig:results:shor:nosv}
  \end{minipage}
  \begin{minipage}[b]{0.48\textwidth}
  \includegraphics[width=\columnwidth]{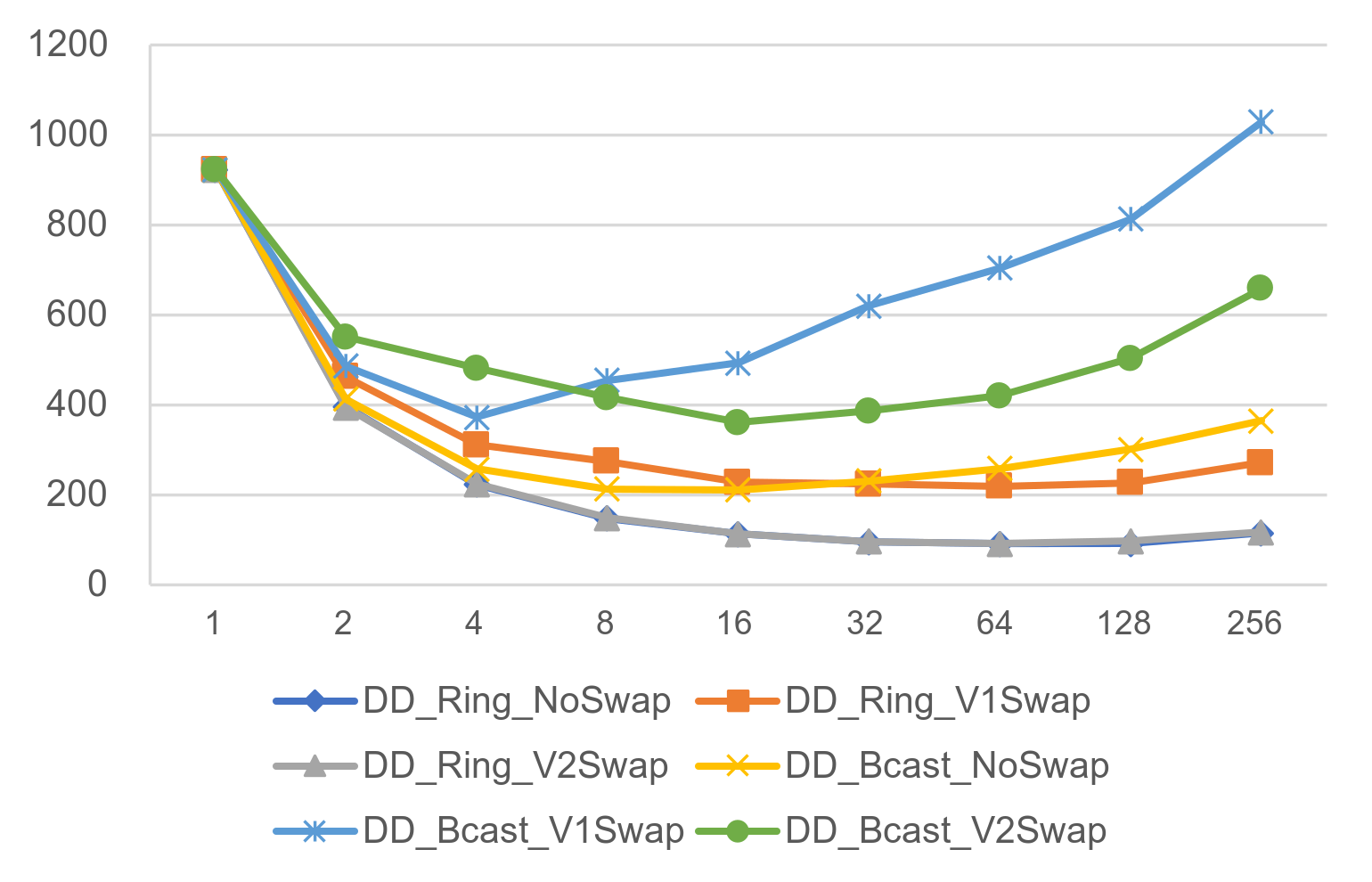}
  \caption{QCBM: nNodes and run-time (sec) without statevector-based results}
  \label{fig:results:qcbm:nosv}
  \end{minipage}
\end{figure*}

To help understand the relationship between the number of nodes and the relative execution time, a line graph is shown in Fig. \ref{fig:results:qcbm}.
The graphs for the statevector-based simulator use 29 qubit results. Fig. \ref{fig:results:qcbm:nosv} is the line graph without statevector-based results, and its vertical axis is not logged.

\begin{table*}[t]
\centering
\caption{QCBM (random circuit) Experimental Results (DD:Decision Diagram-based, SV: Statevector-based)}
\label{tab:qcbm}
\begin{tabular}{@{}crrccrrrrrrrrr@{}}
\toprule
\multicolumn{1}{c}{\multirow{2}{*}{Simulator}} & \multicolumn{1}{c}{\multirow{2}{*}{nQubits}} & \multicolumn{1}{c}{\multirow{2}{*}{nGates}} & \multicolumn{1}{c}{\multirow{2}{*}{MPI}} & \multicolumn{1}{c}{\multirow{2}{*}{SWAP}} & \multicolumn{9}{c}{nNodes \& run-time (sec)} \\ \cmidrule(l){6-14} 
\multicolumn{1}{c}{} & \multicolumn{1}{c}{} & \multicolumn{1}{c}{} & \multicolumn{1}{c}{} & \multicolumn{1}{c}{} & 1 & 2 & 4 & 8 & 16 & 32 & 64 & 128 & 256 \\ \midrule
DD & 20 & 761 & Ring & No & 924 & 396 & 223 & 148 & 113 & 95 & 91 & 91 & 115 \\
DD & 20 & 761 & Ring & v1 & 924 & 463 & 311 & 275 & 228 & 224 & 219 & 227 & 271 \\
DD & 20 & 761 & Ring & v2 & 924 & 395 & 224 & 148 & 113 & 96 & 91 & 97 & 117 \\
DD & 20 & 761 & Bcast & No & 924 & 413 & 257 & 213 & 211 & 231 & 258 & 301 & 365 \\
DD & 20 & 761 & Bcast & v1 & 924 & 487 & 372 & 455 & 494 & 621 & 704 & 813 & 1,030 \\
DD & 20 & 761 & Bcast & v2 & 924 & 551 & 481 & 417 & 360 & 387 & 420 & 503 & 659 \\
SV & 20 & 761 & \cite{imamura2022mpiqulacs} & No & 2.74 & 2.44 & 2.38 & 2.32 & 2.29 & 2.38 & 2.30 & 2.33 & 2.86\\
SV & 20 & 761 & \cite{imamura2022mpiqulacs} & Yes & 2.90 & 2.55 & 2.41 & 2.36 & 2.34 & 2.28 & 2.33 & 2.43 & 2.33\\
SV & 29 & 1,103 & \cite{imamura2022mpiqulacs} & No & 378 & 207 & 116 & 64 & 39 & 23 & 16 & 13 & 9 \\
SV & 29 & 1,103 & \cite{imamura2022mpiqulacs} & Yes & 378 & 194 & 104 & 56 & 30 & 17 & 11 & 7 & 6 \\ \bottomrule
\end{tabular}
\end{table*}

\section{Evaluation}\label{sec:evaluation}
\subsection{Evaluation of ring communication}
This section evaluates the effects of the ring communication method in Sec. \ref{sec:methods:comm} in comparison with the communication method using broadcast.

In the case of Shor algorithm, it was found that the ring communication method performed faster than the broadcast method.
With the fastest automatic swap insertion method, v1, the ring communication resulted in a speedup of up to 35\%. Moreover, the larger the number of nodes, the greater the effect is.

In the case of the QCBM circuit, it was found that the ring communication works faster than the broadcast communication regardless of the swap insertion methods.
However, 32 and 64 nodes had the fastest run times, while 128 and 256 nodes were slower. Thus, the run-time can be rather slowed by increasing the number of nodes, because there is an overhead of inter-node communication even when ring communication is used.

The actual connection between nodes is possibly the reason why the ring communication is faster than the broadcast communication.
A typical parallel computer supports communication from any node to any node but those connections are not necessarily direct ones. This is why the overhead of broadcast communication tends to be large.
Furthermore, the ring communication can easily be realized with direct physical connections. The experimental environment uses Tofu InterConnect-D, which is a six-dimensional mesh/torus, and most of the communication between adjacent nodes can be obtained by a direct physical connection.

\begin{itembox}[l]{Ring communication}
Ring communication is superior for multi-node decision diagram-based quantum simulators.
\end{itembox}

\subsection{Evaluation of auto swap Insertion}
This section compares the effects of the automatic swap insertion described in Sec. \ref{sec:methods:swap}.

In the case of the Shor algorithm, it performed the fastest when using v1. In particular, the more nodes, the more effective it was: 256 nodes were approximately 60\% faster than v2 and approximately 10 times faster than the case of no swapping. Therefore, choosing the correct auto-swap gate insertion method is important for the Shor algorithm.
For QCBM circuits, an execution time with v2 is nearly equal to the one without auto-swap insertion. v1 is two times slower than v2.

The possible reasons why the Shor algorithm is faster, particularly with v1, are:
First, the Shor algorithm has multiple quantum gates applied to a particular qubit in succession, making swap gate insertion highly effective.
Second, the size of the decision diagram tends to increase when qubits not related to IQFT are at the beginning of the qubit order. Therefore, by using v1, which preserves the original qubit order as much as possible, the decision diagram size is kept relatively small, speeding up calculations.

A possible reason why QCBM is the fastest without the swap gate is its circuit configuration.
Because QCBM has quantum gates applied to every bit one by one and there are fewer consecutive gates applied to a specific qubit, it makes auto-swap insertion less effective.

\begin{itembox}[l]{Auto swap insertion}
When a decision diagram-based quantum simulator is multi-noded, the optimal auto-swap insertion method relies on the target quantum algorithms to be simulated. In some cases, run-time is faster without it.
Under the assumption that the nature of the problem does not rely on its size, researchers should try small problems and find the best choice before tackling larger problems.
\end{itembox}

\subsection{Comparison with the statevector-based simulator (Qulacs)}
The statevector- and decision diagram-based quantum simulators have different algorithms that result in faster run times. In this study, the decision diagram-based simulator ran faster with the Shor algorithm, and the statevector-based simulator ran faster with the QCBM circuit. Therefore, we should first choose the type of quantum simulator according to the target quantum algorithms to be simulated.

If the two simulators run in a similar run time with a single node, it does not matter which one to use. However, if we have a large number of nodes available, the statevector-based simulator tends to provide consistent speedups, as shown in Fig. \ref{fig:results:shor} and \ref{fig:results:qcbm}. Therefore, the statevector-based may be used if memory usage permits, although decision diagram-based simulator may not need large amount of memory depending on quantum circuits.

On the other hand, the results of the 2-node Shor experiment should be noted. The decision diagram-based simulator is about three times faster than the one-node experimental results.
Although a 3-fold speed-up with 2-fold computational resources seems unnatural, the decision diagram-based simulator is realized by combining unique table and cache table, and it is possible that the search speed, etc. can be significantly improved by using multiple nodes. In addition, garbage collection is used for memory management, and the run times may be changed significantly depending on the frequency of garbage collection.

\begin{itembox}[l]{Statevector-based vs. Decision diagram-based}
Different simulators are good at different algorithms. Under the assumption that the nature of the problem does not depend on its size, researchers should try small problems first to find the best choice and then work on larger problems. If single-node run times between statevector-based and decision diagram-based simulators are similar, statevector-based should be used with multi-node environments. However, with a small number of nodes, decision diagram-based quantum simulator may have a higher speedup effect.
\end{itembox}

\section{Conclusion \& Future Works}\label{sec:summary}
This study focuses on the multi-node decision diagram-based quantum simulator and proposes two methods for speeding up: ring communication and automatic swap insertion.
Ring communication is a method in which nodes are connected in a ring and communicate only with neighboring nodes. Although automatic swap insertion has been proposed in the previous studies, we proposed two new automatic insertion methods specific to the decision diagram-based quantum simulator.
We implemented these techniques and evaluated them with the Shor algorithm and QCBM (random) circuits. Consequently, it was found that the ring communication was approximately 35\% faster than the communication using broadcast. In addition, it became clear that the auto-swap insertion should be chosen depending on the algorithm. With the appropriate choices, Shor circuits that need 38 qubits can finish simulation in 147 seconds with 256 nodes, which is 26 times faster than the single node result. As far as the authors' knowledge, this result is the fastest in the world \cite{ddsim,cryptoeprint:2023/092}.

Future research directions include hybrid parallelization. In our implementation, the operation in each node is single-threaded, but multi-threading may increase the simulation speed. Research on multi-threading has proposed a parallel processing mechanism using worker threads and a multi-threaded table design~\cite{qsw2023}, which may be incorporated into the proposed methods.

We used the Shor algorithm and the QCBM circuit in the experiments, and we would like to investigate the performance with other quantum algorithms. Moreover, large circuits which cannot be simulated single node may be able to be simulated with multi-nodes using larger amount of overall memory. We would like to investigate the use of multi-node based simulators
to simulate larger circuits.

\bibliography{main}
\bibliographystyle{IEEEtran}
\end{document}